\documentclass[12pt,preprint]{aastex}

\newcommand{\kev}{keV}
\newcommand{\fe}{Fe~K$\alpha$}
\newcommand{\etal}{et al.}

\newcommand{\cyg}{Cyg~X-1}
\newcommand{\ele}{$\mathcal{N}_E$}
\newcommand{\flx}{cm$^{-2}$~s$^{-1}$~keV$^{-1}$}

\usepackage{graphicx}

\begin{document}

\title{The Contribution of Particle Impact to the Production of
Fe~K$\alpha$ Emission from Accreting Black Holes}


\author{D. R. Ballantyne}
\affil{Canadian Institute for Theoretical Astrophysics, McLennan Labs,
60 St. George Street, Toronto, Ontario, Canada M5S 3H8}
\email{ballantyne@cita.utoronto.ca}

\and

\author{A. C. Fabian}
\affil{Institute of Astronomy, University of Cambridge, Cambridge,
U.K. CB3~0HA}

\begin{abstract}
The iron K$\alpha$ line is perhaps the most important spectral
diagnostic available in the study of accreting black holes. The line
is thought to result from the reprocessing of external X-rays by the
surface of the accretion disk. However, as is observed in the solar
corona, illumination by energetic particles may also produce line
emission. In principle, such a process may be uncorrelated with the
observed X-rays and could explain some of the unexpected variability
behavior of the \fe\ line. This paper compares predictions of iron
K$\alpha$ flux generated by impacting electrons and protons to that
from photoionization. Non-thermal power-laws of electrons are
considered as well as thermal distributions of electrons and
virialized protons. The electrons are thought to originate in a
magnetically dominated accretion disk corona, while the protons are
considered in the context of a two phase (hot/cold) accretion
scenario. In each case, the \fe\ flux from particle impact is found to
be $< 1$\% of that produced by photoionization by a hard X-ray
power-law (normalized to the same energy flux as the particles). Thus,
the electrons or protons must strike the disk with $10^2$--$10^4$
times more energy flux than radiation for particle impact to be a
significant producer of \fe\ flux. This situation is difficult to
reconcile with the observations of hard X-ray spectra, or the proposed
particle acceleration mechanisms in the accretion disk
corona. Truncated accretion flows must be externally illuminated by
hard X-rays in order to produce the \fe\ line, as proton impact is
very inefficient in generating line emission. In contrast to the Sun,
our conclusion is that, with the possible exception for localized
regions around magnetic footpoints, particle impact will not be an
important contributor to the X-ray emission in accreting black holes.
\end{abstract}



\section{Introduction}
\label{sect:intro}
An iron K$\alpha$ emission line at 6.4~\kev\ is commonly observed in
the X-ray spectra of active galactic nuclei (AGN) and Galactic black
hole candidates (GBHCs). The equivalent widths (EWs) of the lines are
usually $\sim$100--300~eV \citep*{mdp93,nan97,rt00}, suggesting that a
significant column of iron is necessary to produce the required
emission. This was confirmed with the discovery of the Compton
reflection hump between 20--30~\kev\ \citep{pou90,np94} which is a
natural outcome of X-ray reprocessing in optically-thick material
\citep{lw88,gr88}. In some cases, sensitive observations of the \fe\
line have found them to possess broad low-energy wings, consistent
with the profile expected from a spinning accretion disk close to the
black hole (see the reviews by \citet{fab00} and
\citealt{rn03}). These relativistically broadened lines have been seen
in both AGN \citep[e.g.,][]{tan95,nan97,nan99,fab02} and GBHCs
\citep[e.g.,][]{mar02,mil02a,mil02b,mil02c}, and provide strong
evidence that the optically-thick reflector responsible for the shape
of the X-ray continuum is the inner region of the accretion
disk. Thus, there has been considerable attention paid in recent years
to the \fe\ line and X-ray reflection features in accreting black
holes, as they may probe the physics of the flows only a few
Schwarzschild radii from the event horizon.

The phenomenology outlined above has given rise to a theoretical
framework in which the X-rays are generated by some process in a
corona above the accretion disk
\citep*{gal79,haa91,haa93,haa94}. Roughly half of the X-ray power
illuminates the disk, is reprocessed, and then reemitted as a
reflection spectrum which carries the \fe\ line and reflection hump
\citep*{gf91,mpp91}. The iron line is produced by fluorescence; that
is, incoming X-rays (usually in the form of a power-law) with energies
greater than 7.1~\kev\ can remove a K shell electron from an iron atom
or ion. The excited ion can then relax by a radiative transition into
the K shell, or by the expulsion of an outer shell electron (the
non-radiative Auger effect). With this mechanism the flux in the line
is directly proportional to the ionizing flux \citep{bas78,bai79}, so
that changes in the illuminating continuum should give rise to
correlated changes in the Fe line flux. However, long monitoring
campaigns of certain bright AGNs have shown that, at least in those
sources, the \fe\ line exhibits weak variability on short timescales
despite large changes in the continuum
\citep[e.g.,][]{chi00,rey00,ve01}. Revisions to the standard
disk-corona model are necessary to explain this behavior
\citep*{rey01,fv03,bal03}.

Fluorescence from photoionization is not the only mechanism able to
generate an iron K line. Electron bombardment of the accretion disk
can also produce line emission from iron if the kinetic energy of the
particle is greater than 7.1~\kev. This process has been used to
explain enhanced line emission from the Sun \citep[e.g.,][]{zds92},
where electrons energized in the solar corona travel down magnetic
field lines and strike the photosphere \citep{ds89}.  As accretion disks likely
also possess such magnetized coronae \citep{mf01}, it is possible that
similar situations would arise and result in \fe\ emission by
electron bombardment.
 
In accretion disk coronae there exists a bath of thermal electrons
with temperature $kT \approx 100$--$200$~\kev\ (which generates the
primary X-ray power-law through Comptonization of the disk UV
photons). There may also be a non-negligible amount of
non-thermal electrons present in the corona. Evidence for these
particles has been found in \textit{Compton Gamma-ray Observatory}
observations of \cyg\ in its soft state \citep{gie99} and GRS~1915+105
\citep{zdz01}. These data, when combined with simultaneous hard X-ray
observations, show that the X-ray power-law extends to energies $>
600$~\kev. Such spectra can be well fit by a model of Comptonization
in a hybrid thermal/non-thermal plasma \citep{cop99}. At present there
are no analogous observations for radio-quiet AGN, but it is expected
that the disk-corona physics will be similar for the two systems.

Proton illumination will also produce \fe\ emission in a manner
analogous to electron bombardment.  This may occur in a geometry where
the optically thick accretion disk is disrupted into a hot ion flow
close to the black hole
\citep*[e.g.,][]{mm94,rc00,man00,mk00,mlm00,sd02}. This configuration
may be relevant to GBHCs in the low/hard state
\citep*{emn97,dz99,bdn03} and low luminosity AGNs
\citep[e.g.,][]{qua99}. In this scenario, the accretion energy is
envisaged to be dissipated into the protons within the central
optically thin flow, which then rapidly reach the virial temperature
as they can be thermally decoupled from the electrons
\citep*{sle76}. The majority of the accretion energy is carried into
(or advected) into the black hole by the protons with little or no
observational signatures. Thus, these ``advection dominated accretion
flows'' (ADAFs) can be very radiatively inefficient
\citep*{ich77,ree82,ny94}. However, if the optically thick disk
partially overlaps with the ADAF (for example, in the region where
evaporation of the disk is taking place), then the hot protons in the
ADAF will interact with the disk, significantly heating its surface
\citep*{sh00,ds00,dds02} and possibly producing \fe\ emission. Proton
impact ionization would also be relevant if a two-temperature corona exists above an
untruncated accretion disk \citep{dbf97}. 

If energetic coronal electrons (in an accretion disk-corona model) or
virialized protons (in a truncated disk-ADAF model) strike the
accretion disk they would produce \fe\ emission. If this process is
sufficiently efficient it could generate an observable line that is
not strongly correlated with the continuum \citep[e.g.,][]{pet02}.
This paper considers \fe\ production from particle impact onto
accretion disks and compares it with the more traditional mechanism of
photoionization. We consider incident beams consisting of thermal or
non-thermal electrons and virialized protons. The goal is to evaluate
under which circumstances, if any, particle bombardment will be an
important contributor to the \fe\ line.  The theory needed to
calculate the expected line flux is outlined in the next section, and
Section~\ref{sect:res} presents the results. In
Section~\ref{sect:discuss} we consider the implications of the
computations on particle acceleration mechanisms and the two different
models of accretion geometry. A summary and our conclusions are presented
in Section~\ref{sect:summ}.

\section{Mechanisms of Fe~K$\alpha$ Production}
\label{sect:theory}

\subsection{Electron Bombardment}
\label{sub:electrons}
Consider a beam of electrons incident on the normal of an accretion
disk with a spectral flux of \ele\ electrons~\flx\ at kinetic energy $E$. In
both AGN and GBHCs the disk is hot enough so that both hydrogen and
helium are fully ionized. Therefore, each incident electron will lose
energy by Coulomb interactions with the hydrogen ions as it travels
through the disk \citep{ems78}:
\begin{equation}
{dE \over dt} = -2\pi e^4 \Lambda {n v \over E},
\label{eq:e1}
\end{equation}
where $e$ is the electron charge in e.s.u., and $n$, $v$ and $E$ are
the instantaneous total hydrogen number density, electron velocity and
energy, respectively. $\Lambda$ is the Coulomb logarithm
\begin{equation}
\Lambda = \ln \left({E b_0 \over e^2} \right),
\label{eq:coulog}
\end{equation}
where $b_0$ is the maximum impact parameter for the scattering
event. Equation~\ref{eq:e1} differs from the classic analysis by \citet{spi62}
by ignoring the the correction due to thermal motion of the target electrons
\citep{ems78}. This ``cold target'' approximation holds if $E \gg kT$,
where $T$ is the electron temperature of the accretion disk. Since we
are interested in incoming electrons with $E > \chi$, where
$\chi=7.112$~\kev\ is the ionization potential for neutral iron, this
condition will hold for both AGN ($kT \sim 10$~eV) and GBHC ($kT \sim
500$~eV) disks. 

As each incident electron decelerates\footnote{About 10$^{-5}$ of the
energy lost by the electron through scattering is radiated away as
bremsstrahlung emission in the X-ray band, with the remainder ending
up as heat \citep[see, e.g.,][]{tat02}. This radiation is ignored
here.} from its initial energy $E_0$ it can eject a K shell electron
from iron as long as it energy remains above $\chi$. Thus, the
resulting \fe\ line flux in photons~cm$^{-2}$~s$^{-1}$ is
\citep[e.g.,][]{epd86}
\begin{equation}
\mathcal{F}_{ei} = {1 \over 2} \beta \omega \int_{0}^{\infty}
\mathcal{N}_E(E_0) \left[ \int_{E=E_0}^{E=\chi} \sigma_{ei}(E)
n_{\mathrm{Fe}} v(E) dt \right] dE_0,
\label{eq:fl1}
\end{equation}
where $\omega$ is the K fluorescence yield of iron, $\beta$ is the
K$\alpha$ to K$\beta$ branching ratio,
$\sigma_{ei}$ is the K-shell ionization cross-section of iron due
to electron impact, and $n_\mathrm{Fe}$ is the number density of iron
atoms. The inner integral in the above equation counts the number of
\fe\ photons produced by an electron with initial energy $E_0$ as it
loses energy passing through the disk. This then must be integrated
over the distribution of initial electron energies to find the total
\fe\ line flux. The factor of one-half in front of the expression
accounts for the fact that roughly half of the K$\alpha$ photons
produced will be emitted away from the disk surface. We assume here,
and in the treatment for photoionization described below, that all of the
line photons emitted toward the surface escape to infinity.

Equation~\ref{eq:e1} can be used to rewrite the above expression in a
more useful form:
\begin{equation}
\mathcal{F}_{ei} = {1 \over 2} {\beta \omega A_{\mathrm{Fe}} \over 2
\pi e^4 \Lambda} \int_{\chi}^{\infty}
\mathcal{N}_E(E_0) \left[ \int_{\chi}^{E_0} \sigma_{ei}(E)
E dE \right] dE_0,
\label{eq:flx-ei}
\end{equation}
where $A_{\mathrm{Fe}}$ is the abundance of iron relative to hydrogen
in the disk, and we have changed the lower limit of the outer integral
to the iron K-shell ionization energy (lower energies cannot produce
K$\alpha$ lines). Note that this expression is independent of the
density distribution of the target.

Of course, electrons with larger energies travel a greater distance
before their kinetic energy falls below $\chi$. \citet{ems78} showed
that an electron with initial energy $E_0$ will have an energy
$E=E_0(1-6\pi e^4 \Lambda N_{\mathrm{H}}/E_0^2)^{1/3}$ after
traversing a hydrogen column $N_{\mathrm{H}}$. Therefore, the minimum
initial energy needed by an electron to just be able to K-shell ionize
an iron atom after passing through a column $N_{\mathrm{H}}$ is the
solution of
\begin{equation}
E^3_{\mathrm{min}} - 6\pi e^4 \Lambda N_{\mathrm{H}}E_{\mathrm{min}} -
\chi^3=0.
\label{eq:emin}
\end{equation} 
Using this relation to replace the infinity in the outer integral of
equation~\ref{eq:flx-ei}, will allow us to calculate
$\mathcal{F}_{ei}$ as a function of $N_{\mathrm{H}}$ into the
accretion disk.

\subsection{Proton Bombardment}
\label{sub:protons}
If protons are the incident particles then the equation for the energy
loss rate is modified only\footnote{Strictly speaking, $\Lambda$
is a function of the reduced mass of the scattering
interaction and so will change with the different incident particle. However,
the dependence is logarithmic in nature and so can safely be
neglected.} by a factor of $m_p/m_e$ \citep{ems78}:
\begin{equation}
{dE \over dt} = -2\pi e^4 {m_p \over m_e}\Lambda {n v \over E},
\label{eq:p1}
\end{equation}
where $m_p$ is the proton mass, $m_e$ is the electron mass, and now
$E$ is the kinetic energy of the impacting proton. Note that protons
lose their energy much more rapidly than electrons. As before, we have
assumed the ``cold target'' approximation which in this case requires
$E \gg (m_p/m_e)kT$. For AGN disks, this condition implies $E \gg
18$~keV, while it is $E \gg 900$~keV for the hotter GBHC disks. The
virial temperature for a proton a distance $r$ away from a black
hole is $kT_{\mathrm{vir}}=156 (r/R_{\mathrm{S}})^{-1}$~MeV, where
$R_{\mathrm{S}}=2GM/c^2$ is the Schwarzschild radius for a black hole
of mass $M$. Therefore, assuming the protons in the ADAF reach the
virial temperature, their energies will be greater than 1~MeV if $r <
100$~$R_{\mathrm{S}}$ and so equation~\ref{eq:p1} will be valid for
both AGN and GBHCs.

We can now write down the analogous expression to
equation~\ref{eq:flx-ei} for the \fe\ line flux due to proton impact:
\begin{equation}
\mathcal{F}_{pi} = {1 \over 2} {\beta \omega A_{\mathrm{Fe}} \over 2
\pi e^4 \Lambda} {m_e \over m_p} \int_{\chi}^{\infty}
\mathcal{P}_E(E_0) \left[ \int_{\chi}^{E_0} \sigma_{pi}(E) E dE
\right] dE_0,
\label{eq:flx-pi}
\end{equation}
where $\mathcal{P}_E$ is the spectral proton flux (protons~\flx)
normally incident on the disk, and $\sigma_{pi}(E)$ is the K-shell
ionization cross-section of iron due to a proton with kinetic energy
$E$. This expression differs notably from equation~\ref{eq:flx-ei} by
a factor of $m_e/m_p$, therefore, unless this is compensated by the
cross-section, protons will be a much less significant producer of
\fe\ photons than electrons.

After passing through a column $N_{\mathrm{H}}$, a proton with initial
energy $E_0$ will have an energy \citep{ems78} 
\begin{equation}
E=E_0 \left(1-4\pi e^4 \Lambda \left( {m_p \over m_e} \right)
  {N_{\mathrm{H}} \over E_0^2} \right)^{1/2}.
\end{equation}
Therefore, as with the electrons (eq.~\ref{eq:emin}), we can calculate
the minimum kinetic energy a proton must have to just be able to
ionize iron at the bottom of a column $N_{\mathrm{H}}$:
\begin{equation}
E_{\mathrm{min}} = \sqrt{\chi^2 +4 \pi e^4 \Lambda (m_p/m_e)
  N_{\mathrm{H}}}.
\label{eq:emin-p}
\end{equation}
Replacing the infinity with $E_{\mathrm{min}}(N_{\mathrm{H}})$ in the
outer integral in equation~\ref{eq:flx-pi} gives the total \fe\ flux
produced through a column $N_{\mathrm{H}}$.

\subsection{Photoionization}
\label{sub:photons}
There are many previous studies which have computed the \fe\ fluxes
expected from photoionization \citep{hw77,bas78,bas79,bai79}. In this work,
however, we will use the simplest expression for the line flux:
\begin{equation}
\mathcal{F}_{\mathrm{photo}}= {1 \over 2} N_{\mathrm{H}} A_{\mathrm{Fe}} \omega
\beta \int_{\chi}^{\infty} F_{\epsilon}(\epsilon)
\sigma_{\mathrm{photo}}(\epsilon) d\epsilon,
\label{eq:flx-fl}
\end{equation}
where $F_{\epsilon}(\epsilon)$ is the photon flux (ph~\flx) incident
on the normal of the accretion disk at energy $\epsilon$, and
$\sigma_{\mathrm{photo}}$ is the K-shell photoionization cross-section
of iron. As in the expression for electron impact production of \fe\
(eq.~\ref{eq:flx-ei}), this equation assumes that iron is the only
metal in the gas. Furthermore, equation~\ref{eq:flx-fl} is only valid
in the optically-thin limit. That is, we are ignoring any attenuation
of the incident X-rays or the outgoing \fe\ line photons. Therefore,
we are limited to columns $N_{\mathrm{H}} \leq
1/(1.2\sigma_{\mathrm{T}}) = 1.25\times 10^{24}$~cm$^{-2}$, where
$\sigma_{\mathrm{T}}$ is the Thomson cross-section, and we have assumed a 10\%
abundance of helium (by number). The implications of limiting the
calculations to Thomson thin columns are considered in the next
section.

\subsection{Charge Exchange}
\label{sub:charge}
Another particle process that can produce \fe\ emission is charge
exchange collisions. These are reactions of the following form:
$\mathrm{Fe}^{q+} + \mathrm{A} \rightarrow \mathrm{Fe}^{(q-1)+} +
\mathrm{A}^+$, where A is usually H, H$_{2}$, or He. The product Fe
ion is often left in an excited state and will produce line emission
as it relaxes. This collision differs from those described above in
that the Fe ion is the energetic particle and the hydrogen or helium atom
is the target. This process may be important for explaining the strong
and broad K$\alpha$ lines from hydrogenic and helium-like Fe observed
in the Galactic ridge X-ray emission \citep*{tmh99,tan02}. In this case, highly
ionized Fe ions in the cosmic-ray distribution interacting with
neutral Galactic gas would produce the needed line emission.

For the accretion disk scenario that is considered here, it is
unlikely that charge exchange will be an important process, as there
are very few neutral H and He atoms available for the reaction. Other
metal ions could act as donors in a collision with Fe, but, for
typical cosmic abundances, encounters with the correct kinematics
would be extremely rare. However, charge exchange may possibly be
relevant to iron K lines emitted from the molecular torus of AGN
unification schemes \citep*[e.g.,][]{ghm94,kmz94} if there is a flux
of energetic Fe ions incident on this gas.

\section{Results}
\label{sect:res}

\subsection{Electron Impact Versus Photoionization}
\label{sub:eres}
To determine how efficient electron impact is in producing an \fe\
line in comparison to photoionization, we
calculate the ratio of equations~\ref{eq:flx-ei} and \ref{eq:flx-fl}:
\begin{equation}
{\mathcal{F}_{ei} \over \mathcal{F}_{\mathrm{photo}}} =
{\int_{\chi}^{E_{\mathrm{min}}(N_{\mathrm{H}})}
\mathcal{N}_E(E_0) \left[ \int_{\chi}^{E_0} \sigma_{ei}(E)
E dE \right] dE_0 \over 2 \pi e^4 \Lambda N_{\mathrm{H}}
\int_{\chi}^{\infty} F_{\epsilon}(\epsilon) \sigma_{\mathrm{photo}}(\epsilon)
d\epsilon}.
\label{eq:ratio}
\end{equation}
Considering this ratio will also help negate the effects of absorption
and scattering on the K$\alpha$ line that were ignored in the
individual expressions. This value will nevertheless be only a upper
limit, because in the true Thomson thick case high energy photons can
scatter down to energies where they can ionize iron and produce a
K$\alpha$ photon. Some fraction of the line photons will escape the
layer, but to accurately treat this process requires a numerical
prescription. 

The K-shell photoionization cross-section ($\sigma_{\mathrm{photo}}$)
was taken from the tabulation by \citet{vy95}. The empirical fit
formula of \citet{hom98} was used to compute the cross-section for
K-shell ionization by electron impact\footnote{There is a misprint in
the expression for $G_r$ in the paper by \citet{hom98}, so the form
quoted by \citet{qua76} was used in its place.}. These cross-sections are
plotted between $\chi$ and 200~\kev\ in Figure~\ref{fig:xsect1}, where
the x-axis is used for both photon energy and electron kinetic
energy. $\sigma_{\mathrm{photo}}$ is roughly proportional to
$\epsilon^{-3}$ and so falls off much more rapidly than
$\sigma_{\mathrm{ei}}$; however, $\sigma_{\mathrm{photo}}$ is
significantly greater than $\sigma_{\mathrm{ei}}$ for energies below
40~\kev.
%
%

A power-law photon spectrum was assumed for all the calculations
described below: $F_{\epsilon}(\epsilon) = F_0
(\epsilon/\chi)^{-\Gamma}$, where $\Gamma=1.7, 1.8, 1.9$ or $2.0$ is
the photon index. For each value of $\Gamma$ the spectrum was
normalized so that the total energy flux between $\chi$ and 200~\kev\
at the surface of the disk was 1~\kev~cm$^{-2}$~s$^{-1}$. A value of
25 is chosen for the Coulomb logarithm $\Lambda$ \citep{br71}, but, as
will be seen below, the results are insensitive to its value.

The value of $N_{\mathrm{H}}$ and equation~\ref{eq:emin} determines
the upper-limit to the outer integral in the numerator of
equation~\ref{eq:ratio}. K$\alpha$ photons produced by electrons with
kinetic energies larger than this are
accounted for in a later bin where the column is large enough for
the energy of those electrons to pass below $\chi$. Since the largest
column we consider is $10^{24}$~cm$^{-2}$, there is a slight error as
we do not count any \fe\ emission produced by electrons that can pass
through this column. However, equation~\ref{eq:emin} shows that only
electrons with kinetic energies greater than $\sim 3.5$~MeV can still
ionize iron beyond a column of $10^{24}$~cm$^{-2}$. For the
electron distributions we consider, there is a negligible amount of
electrons with $E > 3.5$~MeV, so this slight error will not impact the
results. 

\subsubsection{Power-law $\mathcal{N}_E$}
\label{subsub:eplaw}
Modeling of the hard X-ray tails in the GBHCs Cyg~X-1 and
GRS~1915+105 have shown that the non-thermal electrons responsible for
this emission can be described as a soft power-law \citep{gie99,zdz01}. Therefore,
we first consider electron distributions of the form $\mathcal{N}_E
\propto (E/\chi)^{-\gamma}$, where $\gamma=1, 2, 3$ or $4$ is the number
  index. As with the photon distribution, the incident electron
  power-laws are normalized such that the total energy flux between
  $\chi$ and $200$~\kev\ is 1~\kev~cm$^{-2}$~s$^{-1}$. In this way, roughly
  the same amount of energy is being deposited into the material by
  both the electrons and photons.

\begin{figure}
\epsscale{1.0}
\plotone{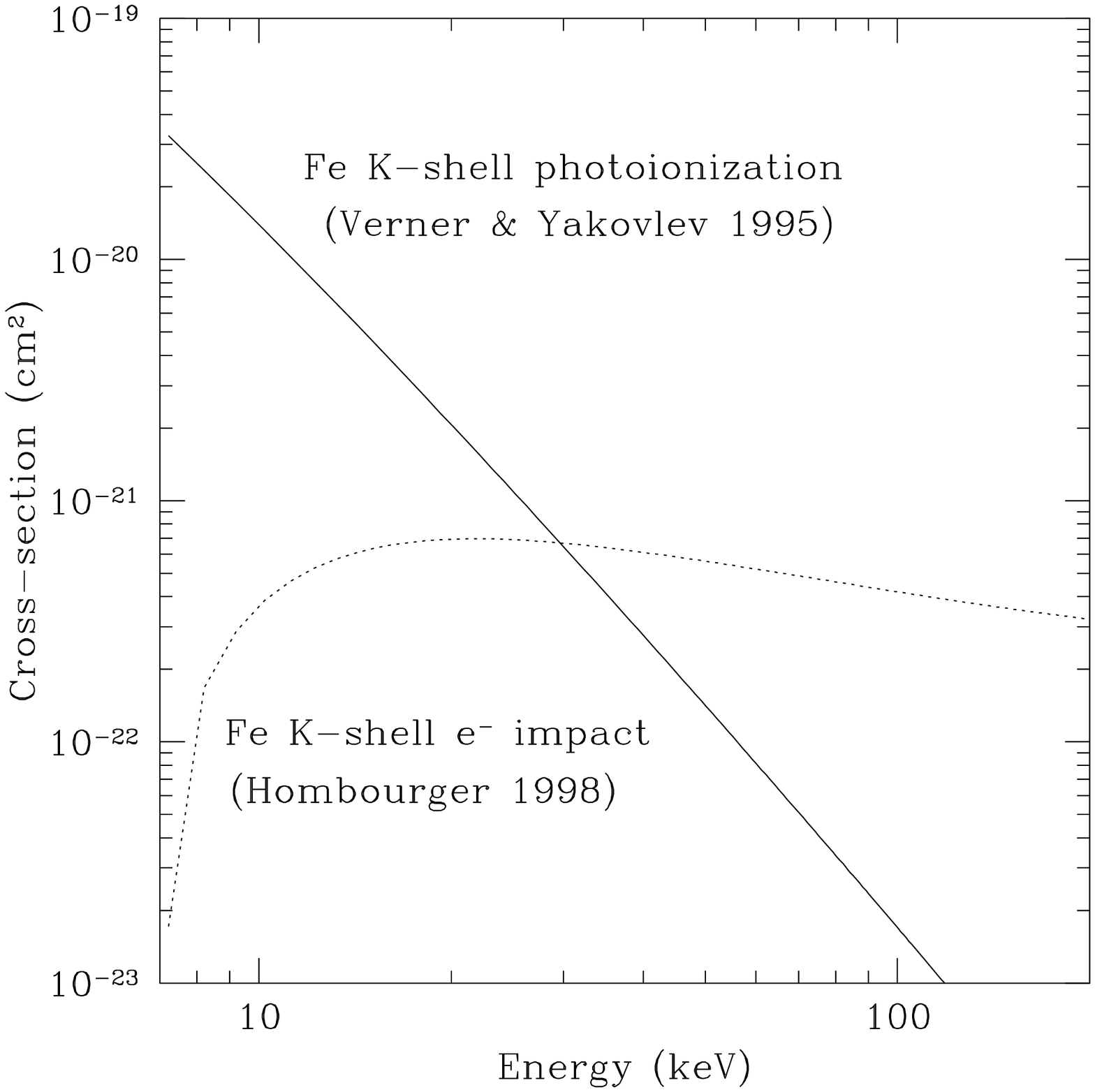}
\caption{K-shell ionization cross-section of iron as a function of
energy for two different ionization processes. The solid line shows
the cross-section for photoionization using the fits of
\citet{vy95}. The dotted line denotes the empirical expression for
electron impact cross-section found by \citet{hom98} (in this case the
x-axis plots electron kinetic energy).}
\label{fig:xsect1}
\end{figure}

\begin{figure}
\epsscale{1.0}
\plotone{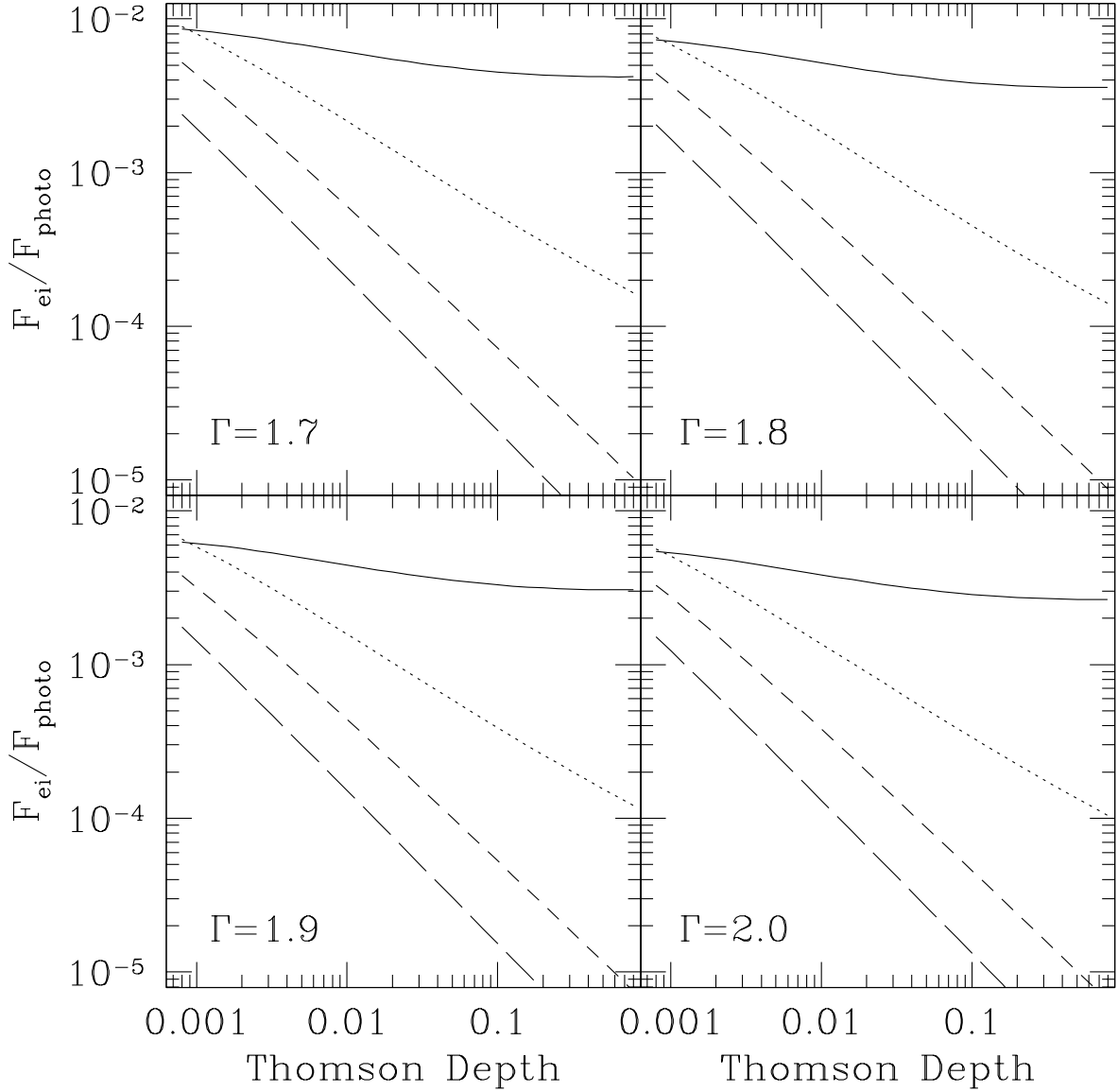}
\caption{Ratio of \fe\ flux from electron impact $\mathcal{F}_{ei}$ to
\fe\ flux from photoionization $\mathcal{F}_{\mathrm{photo}}$. Both incident
spectra were power-laws with the radiation having a photon index
$\Gamma$ and the electrons a number index $\gamma$. Both spectra were
normalized so that the energy flux between $\chi$ and 200~\kev\ was
1~keV~cm$^{-2}$~s$^{-1}$. In each panel
$\mathcal{F}_{ei}/\mathcal{F}_{\mathrm{photo}}$ is plotted for $\gamma=1$ (solid
line), $2$ (dotted line), $3$ (short-dashed line), and $4$
(long-dashed line). In all cases the \fe\ line flux produced by
electron impact is negligible compared to the one produced by
photoionization.}
\label{fig:eplaw}
\end{figure}

%
%
Figure~\ref{fig:eplaw} shows the results of computing the ratio
${\mathcal{F}_{ei}/ \mathcal{F}_{\mathrm{photo}}}$ for the different
values of $\Gamma$ and $\gamma$. Each panel shows the results for an
individual photon index, while the different lines denote the electron
number index. In every case considered here the \fe\ flux from
electron impact is a small fraction of that produced by
photoionization. The maximum ratio found is ${\mathcal{F}_{ei} /
\mathcal{F}_{\mathrm{photo}}} \sim 0.01$ when $\Gamma=1.7$, $\gamma=1$
and the Thomson depth ($\tau_{\mathrm{T}} = 1.2 \sigma_{\mathrm{T}}
N_{\mathrm{H}}$) is $\sim 10^{-3}$. The ratio then drops rapidly as
the column increases as less and less electrons are able to penetrate
further into the gas. The only exception is for $\gamma=1$. This
electron spectrum is very flat so that the energy flux per decade
actually increases with energy, and so there are many high energy
electrons that can still ionize \fe\ at larger column densities. The
electron K-shell cross-section is close to a constant at high energies
(Fig.~\ref{fig:xsect1}), which helps keep the ratio roughly the same
as the column increases. However, even with this extreme spectrum,
electron impact is still an inefficient method to produce \fe\
photons.

As $\gamma$ is increased, the ratio at all $\tau_{\mathrm{T}}$
drops. This is a result of the shape of the cross-section which peaks
at electron energies of $\sim 20$~\kev\ and not at the the ionization
threshold, $\chi$ (Fig.~\ref{fig:xsect1}). Therefore, when the
electron spectrum softens, more electrons have lower
energies, below the peak in the cross-section, and the \fe\ flux
decreases accordingly.

There is also a slight change in ${\mathcal{F}_{ei}/
  \mathcal{F}_{\mathrm{photo}}}$ with the photon-index. The ratio lowers as
  $\Gamma$ increases because for softer photon spectra there are more
  photons at lower energies, which is where $\sigma_{\mathrm{photo}}$
  is larger. This increases the number of K$\alpha$ photons emitted,
  and therefore decreases the ratio.

\subsubsection{Thermal $\mathcal{N}_E$}
\label{subsub:emax}
Observations by \textit{BeppoSAX} have shown that many Seyfert~1
galaxies have high energy cutoffs between 100-300~\kev\ \citep{matt01}. If the hard
X-rays in these AGN are produced by thermal Comptonization, then such
measurements can constrain the temperature of the coronal electrons by
$kT \approx E_c/2$, where $E_c$ is the cutoff energy. We now consider
the production of \fe\ photons by electrons in a thermal distribution,
described by a Maxwell-Boltzmann distribution with $kT = 100, 200,
300$ or $400$~\kev. As before, the electron spectra were normalized so
that the total energy flux incident on the gas between $\chi$ and
200~\kev\ was 1~\kev~cm$^{-2}$~s$^{-1}$, the same as for the photon spectra.

\begin{figure}
\epsscale{1.0} \plotone{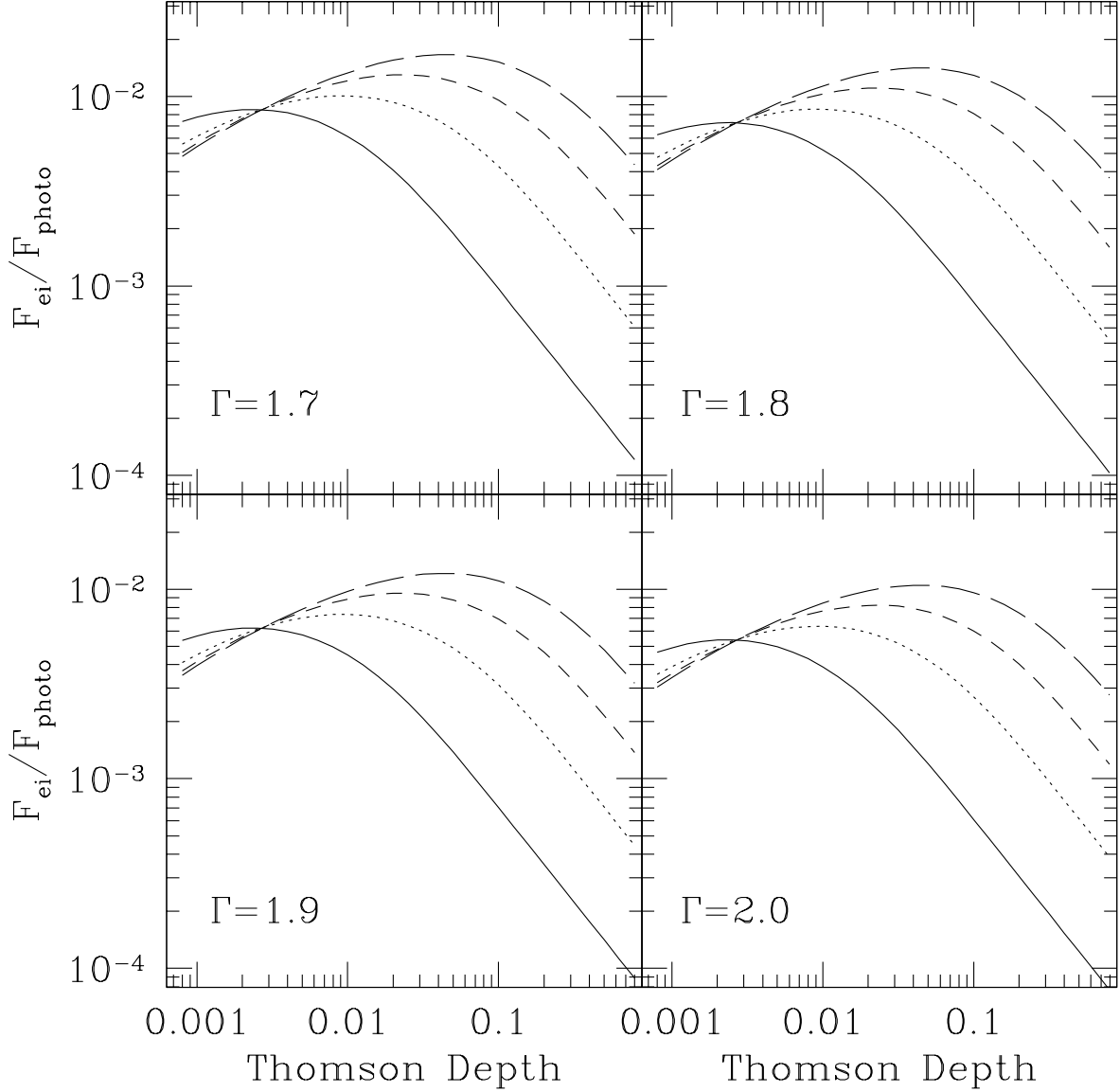}
\caption{Ratio of \fe\ flux from electron impact $\mathcal{F}_{ei}$ to
\fe\ flux from photoionization $\mathcal{F}_{\mathrm{photo}}$. The incident
radiation spectrum was a power-law with a photon index $\Gamma$. The
incident electron spectrum was a Maxwell-Boltzmann distribution of
temperature $kT$. Both spectra were normalized so that the energy flux
between $\chi$ and 200~\kev\ was 1~keV~cm$^{-2}$~s$^{-1}$. In each
panel $\mathcal{F}_{ei}/\mathcal{F}_{\mathrm{photo}}$ is plotted for
$kT=100$~\kev\ (solid line), $200$~\kev\ (dotted line), $300$~\kev\
(short-dashed line), and $400$~\kev\ (long-dashed line). In all cases
the \fe\ line flux produced by electron impact is negligible compared
to the one produced by photoionization.}
\label{fig:emax}
\end{figure}

%
%
Figure~\ref{fig:emax} plots the \fe\ flux ratio between electron
impact and photoionization for electrons in a thermal distribution. 
Again, the flux of K$\alpha$ photons produced by the electrons is
$\la$~1 per cent of that generated by photoionization. An interesting
difference between the thermal and power-law electron distributions is
that the largest flux ratio occurs some distance into the layer for a
Maxwellian population and not at the surface, as was found with the
power-law. This is because the number of electrons in the thermal
distribution peaks at energies $\sim kT$, and can thus ionize Fe to
larger columns. Therefore, the maximum ratio moves to greater
$\tau_{\mathrm{T}}$ for larger $kT$. 

Except for the lowest columns, ${\mathcal{F}_{ei} / \mathcal{F}_{\mathrm{photo}}}$
increases with $kT$. The total number of electrons in the distribution
grows with $kT$, and, since the spectrum is normalized so that the
energy deposited between $\chi$ and 200~\kev\ is the same for each
case, the excess electrons are found at higher energies. This fact,
combined with relative flatness of the K-shell cross-section
(Fig.~\ref{fig:xsect1}), explains the small increase in
${\mathcal{F}_{ei} / \mathcal{F}_{\mathrm{photo}}}$ with $kT$.  

\subsection{Proton Impact Versus Photoionization}
\label{sub:pres}
\begin{figure}
\epsscale{1.0}
\plotone{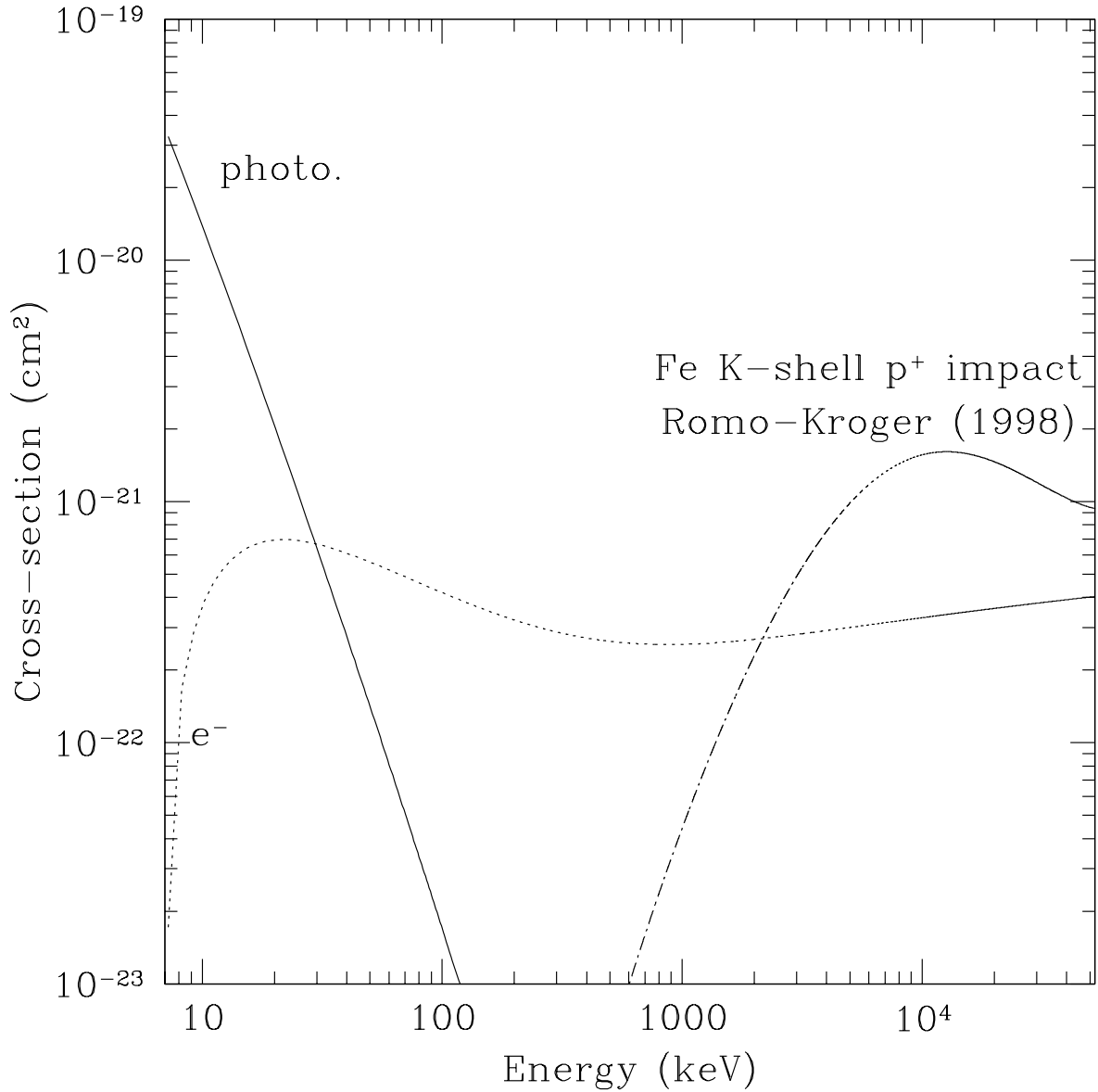}
\caption{Comparison of the Fe K-shell ionization cross-section due to
proton impact (dot-dashed line) with those for electron impact (dotted line) and
photoionization (sold line). The proton impact cross-section was taken
from the empirical fit of \citet{rk98}. This fit is only valid to
proton kinetic energies of $\sim$50~MeV, so the cross-section is
truncated at that point.}
\label{fig:xsect2}
\end{figure}

We form the ratio of \fe\ flux produced by proton impact to that
generated by photoionization by using
equations~\ref{eq:flx-pi} and \ref{eq:flx-fl}:
\begin{equation}
{\mathcal{F}_{pi} \over \mathcal{F}_{\mathrm{photo}}} = {m_e \over m_p}
{\int_{\chi}^{E_{\mathrm{min}}(N_{\mathrm{H}})}
\mathcal{P}_E(E_0) \left[ \int_{\chi}^{E_0} \sigma_{ei}(E)
E dE \right] dE_0 \over 2 \pi e^4 \Lambda N_{\mathrm{H}}
\int_{\chi}^{\infty} F_{\epsilon}(\epsilon) \sigma_{\mathrm{photo}}(\epsilon)
d\epsilon}.
\label{eq:ratio2}
\end{equation}
The calculation of the denominator of equation~\ref{eq:ratio2} is
unchanged from the previous section (e.g., $\Lambda=25$). The K-shell
ionization cross-section for proton impact was computed by using the
empirical formula of \citet{rk98}. This cross-section is plotted along
with the ones for photoionization and electron bombardment in
Figure~\ref{fig:xsect2}. Unfortunately, the formula is only valid for
proton kinetic energies less than $\sim 50$~MeV. Through
equation~\ref{eq:emin-p}, this limits the column density that can be
probed to be $< 3\times 10^{23}$~cm$^{-2}$.
%
%

\subsubsection{Thermal $\mathcal{P}_E$}
\label{subsub:pmax}
This section considers the \fe\ flux produced by a distribution of
protons at the local virial temperature, $kT_{\mathrm{vir}}$. This
situation may be applicable to the case where a relatively cold, thin
accretion disk overlaps with a hot, ADAF-like flow \citep{dds02}. Due to the 50~MeV
cutoff in the proton impact cross-section, we are limited to
proton spectra which do not have many particles above this
energy. A Maxwell-Boltzmann distribution is assumed for the particles,
and three cases are considered: $kT=1.56, 3.12$ and $6.24$~MeV. These
values correspond to the virial temperature at $100$, $50$ and
$25$~$R_{\mathrm{S}}$, respectively. The proton spectra are normalized
so that the total energy flux incident on the gas is
1~\kev~cm$^{-2}$~s$^{-1}$. The \fe\ flux produced by the incident
protons will be compared with that from a power-law photon spectrum,
which is normalized to 1~\kev~cm$^{-2}$~s$^{-1}$ between $\chi$ and
200~\kev.

\begin{figure}
\epsscale{1.0}
\plotone{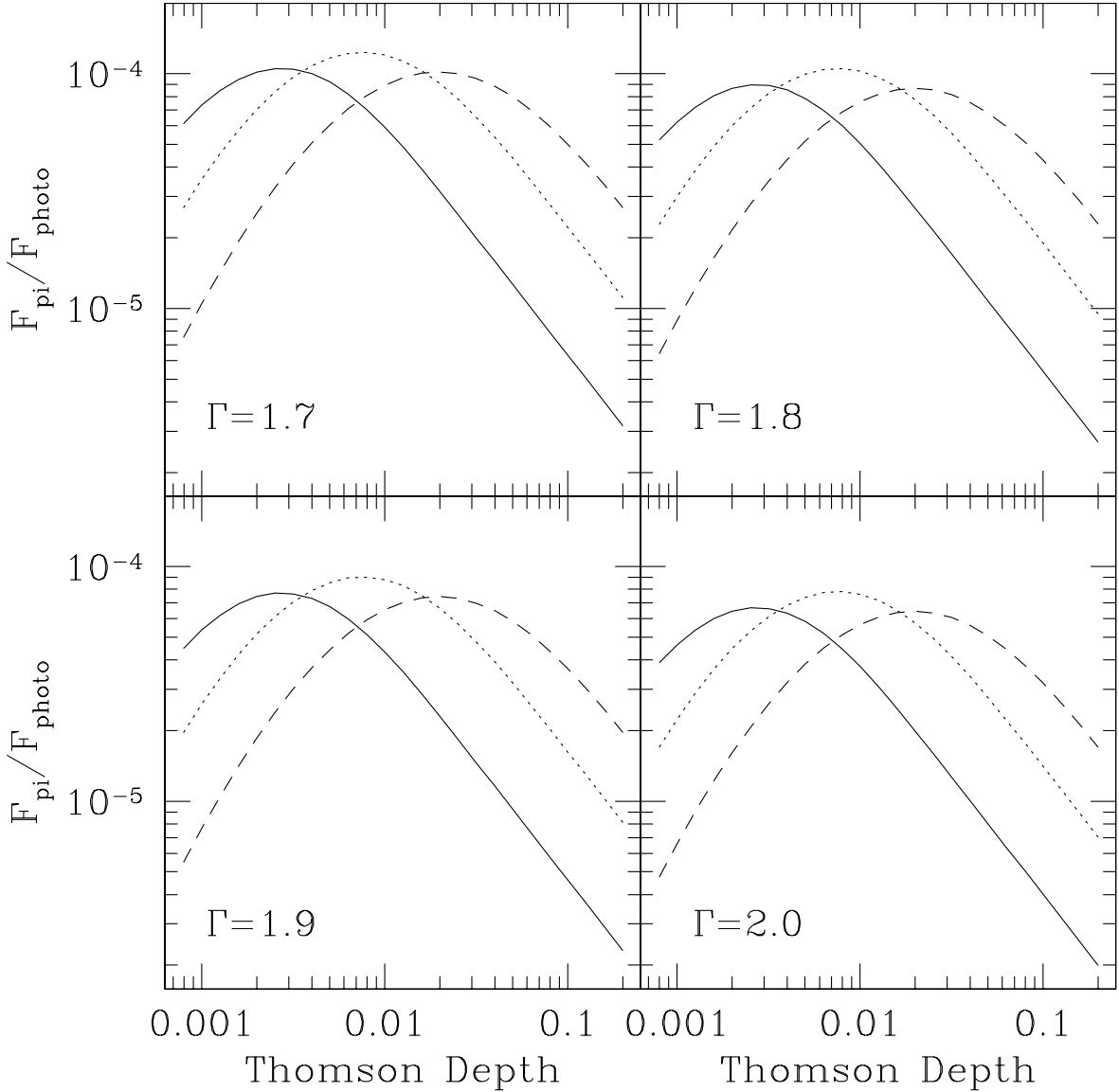}
\caption{Ratio of \fe\ flux from proton impact $\mathcal{F}_{pi}$ to
\fe\ flux from photoionization $\mathcal{F}_{\mathrm{photo}}$. The incident
radiation spectrum was a power-law with a photon index $\Gamma$. The
incident proton spectrum was a Maxwell-Boltzmann distribution of
temperature $kT$. The radiation spectrum was normalized so that the
energy flux between $\chi$ and 200~\kev\ was
1~keV~cm$^{-2}$~s$^{-1}$. The proton spectrum was normalized so that
the total energy flux was 1~keV~cm$^{-2}$~s$^{-1}$. In each panel
$\mathcal{F}_{pi}/\mathcal{F}_{\mathrm{photo}}$ is plotted for $kT=1.56$~MeV
(solid line) corresponding to the virial temperature at
100~$R_{\mathrm{S}}$, $3.12$~MeV (dotted line; 50~$R_{\mathrm{S}}$),
and $6.24$~MeV (short-dashed line; 25~$R_{\mathrm{S}}$). In all cases
the \fe\ line flux produced by proton impact is negligible compared to
the one produced by photoionization. The cutoff in
the proton ionization cross-section (see Fig.~\ref{fig:xsect2}) limits
the calculations to low Thomson depth and low virial temperatures.}
\label{fig:pmax}
\end{figure}

%
%
The results are show in Figure~\ref{fig:pmax}. For each case, it is
found that the protons produce a negligible amount of \fe\ emission as
compared to photoionization. This is in part due to
the normalization conditions. The spectra were normalized so that both
the protons and the photons deposited the same amount of energy (per
unit area, per unit time) into the material. As the majority of the
protons carried 2-3 orders of magnitude more energy than a typical
incident photon, there are far fewer protons in the illuminating
distribution than photons (this should be contrasted with the
situation described in \S~\ref{sub:eres}, where there was roughly equal numbers of
electrons and photons). However, even if the spectra were normalized
to the same number flux, the \fe\ emission produced by protons would
still be much less than that from photoionization.

Of the three values of $kT_{\mathrm{vir}}$ considered, it is the middle
one ($kT_{\mathrm{vir}} = 3.12$~MeV) that results in the largest
${\mathcal{F}_{pi}/ \mathcal{F}_{\mathrm{photo}}}$, due to the peak in its energy
flux residing closest to the maximum cross-section
(Fig.~\ref{fig:xsect2}). On the other hand, it is the
$kT_{\mathrm{vir}}=1.56$~MeV spectrum that has the greatest ratio at
low $\tau_{\mathrm{T}}$. This is because this distribution has a large
number of relatively low energy particles which stop quickly in the
gas, and so produce their K$\alpha$ photons close to the surface. The
other distributions predominantly have much high energy particles
which can penetrate through to larger columns. One other effect that is
noticeable in Fig.~\ref{fig:pmax} is the decrease in
${\mathcal{F}_{pi}/ \mathcal{F}_{\mathrm{photo}}}$ as the photon-index, $\Gamma$,
increases. As explained in Sect.~\ref{subsub:eplaw}, the softer photon
spectra have more photons at low energies, where the cross-section is
larger, and will produce more K$\alpha$ photons, thereby lowering the ratio.

\section{Discussion}
\label{sect:discuss}
We have shown that particle impact is an inefficient method to produce
\fe\ emission as compared to photoionization (see also
\citealt{bas79}). In this section, we attempt to place this result in
the wider context of accretion disk models and particle acceleration
mechanisms. We begin with the electron bombardment results and their
relation to models of coronae above geometrically thin disks. Proton
impact and the two-component disk model is discussed at the end of the
section.

\subsection{Electrons in Accretion Disk Coronae}
\label{sub:discuss-e}
It has been argued for some time that the energy responsible for
heating the electrons in an accretion disk corona is carried and
released by magnetic fields expelled from the disk surface \citep[e.g.][]{gal79,dm98}. The
magnetic energy (ultimately derived from the magnetohydrodynamical (MHD)
turbulence in the body of the disk; \citealt{ms00}) would then be liberated by
reconnection events with neighboring field lines, in a deliberate
analogy to flares in the lower solar corona. Some fraction of this
energy is deposited into electrons which end up in a thermal
distribution. The remaining energy would then be used to accelerate
particles (mostly electrons) to high energies. A key parameter is the
fraction of energy that is funneled into the non-thermal particles for
a given event. Unfortunately, as is discussed below, that number is
not well constrained.

X-ray observations have shown that, in the accretion disk environment,
the newly energized coronal particles give up their energy to the soft
X-ray/UV photons from the accretion disk. This results in the standard
hard X-ray Comptonized power-law that is the hallmark of these
sources. The very hard X-ray tails observed in the spectra of Cyg~X-1
and GRS~1915+105 show that the non-thermal particles are involved in
this process. The results of \S~\ref{sub:eres} have shown that for
electrons to be an important contributor to \fe\ production then $\sim
10^2$--$10^4$ times more energy has to be deposited into the disk by
particles then by the Comptonized photons. For this scenario to be
viable, most of the electrons in the corona would have to not lose
their energy to Comptonization, implying a very low Compton $y$
parameter. However, the observed photon-indices and high energy
cutoffs show that $y \sim 1$ in most of these systems \citep{pet01};
therefore, Comptonization is important. Alternatively, due to outflows
in the corona, the hard X-rays may be beamed away from the disk
resulting in a very small radiation flux on the surface
\citep{bel99}. This scenario is also unlikely, as particles would have
to be both outflowing (to beam the hard X-rays away from the disk) and
streaming toward the disk to generate the \fe\ line. Furthermore, the
Compton reflection hump seen in many Seyfert~1s show that roughly half
the observed X-ray flux must be intercepted by optically thick
material. Perhaps a more fundamental problem with a very strong
particle flux on the disk surface, is that $\gg 99$\% of the incident
energy goes into heat and not radiation \citep{tat02}. This would lead
to a strong expansion of the surface layers and perhaps lead to
evaporation, as is envisaged in the proton-illuminated disks
\citep{dds02}, which would certainly affect the observed
optically-thick signatures such as the Compton-reflection bump.

On the other hand, the non-thermal particles are likely to be tied to
the magnetic field lines in the corona. Thus, there may be a large
particle flux at the position of the magnetic footpoints on the
surface of the accretion disk. In these very local positions, the
electron flux may be much higher than the radiation flux and \fe\ may
be dominated by the electron impact process. Indeed, such bright spots are seen on
the surface of the Sun as sources of hard X-rays \citep{hud94,kun95}. However, such
small regions are unlikely to be detected amidst the unresolved
accretion disk emission.

The dominance of the Comptonized hard X-ray power-law in the spectra
of accreting black holes implies that the acceleration of non-thermal
particles in the corona is not highly efficient. This conclusion is
supported by the above results on \fe\ production by electron
impact\footnote{Analogous arguments can be made against conduction
from the corona as an important heat source for the disk
\citep[e.g.,][]{roz99,rc00b}. In a magnetically dominated corona, the
electrons are more likely to be Compton cooled before conduction can
be effective.}. The mechanism by which electrons are accelerated to
large kinetic energies in accretion disk coronae is not known. Again,
one can look to the solar corona for guidance, where there are three
different processes that are considered to be viable candidates (see
\citet{mill97} and \citet{mill98} for recent reviews). The first is
stochastic acceleration by MHD waves in the coronal plasma which are
excited by the reconnection events. The second is shock acceleration
by the plasma jets produced by reconnection
\citep[e.g.,][]{bf94}. However, it is thought that the shocks are
unlikely to be strong enough to account for most energetic particles
(that is, shocks may only ``pre-accelerate''), although they may
provide the bulk of the heat that goes into the thermal electrons
\citep{dm98}. The third process is direct acceleration by large-scale
electric fields, such as those found in reconnecting current
sheets. The efficiency of all of these processes is, at this point,
largely unknown, but numerical reconnection simulations are beginning
to show that electric field acceleration may be $\la 10$\% efficient
\citep{nod03}.

Before leaving non-thermal electrons, we can use some of the basic
analytical results of electric field acceleration as a consistency
check of our conclusions. When thermal electrons are subjected to an
external electric field they will begin to accelerate along the
direction of the field, but, for most of the particles, this motion
will be limited by Compton drag. However, the collisional drag force
on electrons decreases with the velocity, so that particles with a
large enough speed can ``runaway'' and be accelerated freely by the
electric field. The value of the electric field where this critical
runaway velocity is equal to the thermal velocity $v_e$ of a Maxwellian
distribution is called the Dreicer field and is given by \citep{dre59,dre60}
\begin{equation}
\mathcal{E}_{\mathrm{D}} = e \left ( {\omega_e \over v_e} \right )^2
\Lambda,
\label{eq:dreicer}
\end{equation}
where $\omega_e$ is the electron plasma frequency. In this situation,
the bulk of the electrons in the thermal distribution will be
accelerated. 

In the context of solar flares, \citet{hol85} examined the properties of
accelerated electrons by sub-Dreicer electric
fields\footnote{Super-Dreicer acceleration has been considered by,
e.g., \citet{lit96} and \citet{cl02}, where the acceleration of the electrons are limited by
  transverse magnetic fields.}. Since the magnetic field induced by the
electron beam must be less than the field in the acceleration region ($B$),
\citet{hol85} found that this limited the acceleration rate (per area) to a
maximum of $cB/2\pi eL$, where $L$ is the length of the accelerating current
sheet (see also \citealt*{hkk89}). If the radiation flux on the disk is
$10^{15}$~erg~cm$^{-2}$~s$^{-1}$ between 10~eV and 100~\kev\ \citep*{ros99,brf01},
then, for a $\Gamma=1.9$ spectrum, this translates to $\sim
10^{20}$~erg~cm$^{-2}$~s$^{-1}$ between $\chi$ and 200~\kev. For a
significant \fe\ flux due to electron impact, the energy flux due to
electrons must be $\sim 10^{23}$~erg~cm$^{-2}$~s$^{-1}$
(Fig.~\ref{fig:eplaw}). Converting this to a number flux using a
$\gamma=3$ spectrum, and plugging into the acceleration limit above
gives an estimate of the accelerating magnetic field in the accretion
disk coronae, $B \sim (10^{11} L)$~G. For comparison, the
equipartition magnetic field in a disk is $\sim 10^8$~G \citep{mf01}. Thus,
the assumptions of sub-Dreicer acceleration of non-thermal particles
and significant \fe\ emission from electron impact implies an
extremely strong coronal magnetic field.

\subsection{Protons and Truncated Accretion Disks}
\label{sub:discuss-pr}
Considerable work over the last few years has been done investigating
the properties of disks which change from being cold and geometrically
thin to hot and geometrically thick (see references in
\S~\ref{sect:intro}). One of the greatest difficulties in describing
such a model is the process by which the truncation of the cold disk
occurs. A natural mechanism would be heating the disk until its local
scale height becomes comparable to the radius, effectively evaporating
the inner part of the disk.

Once the two-phase accretion flow is operating, \citet{dds02} showed that 
protons in the hot phase striking the cold disk will be an effective
mechanism for heating the disk surface. Comptonization of the soft
disk photons by this heated layer produced a hard X-ray spectrum not
too dissimilar from what is observed. In \S~\ref{sub:pres} we found
that virialized protons are a highly inefficient way of producing \fe\
emission. Thus, external hard X-ray radiation will need to be included in the
models of truncated accretion flows to explain any Fe~K emission. In the context of the proton
illumination model, it may be possible to determine the relative
importance of radiation heating to proton heating by examining the
equivalent width of the \fe\ line.

\subsection{Other Applications}
\label{sub:apps}
The above sections have shown that it is difficult for particle impact
to be an efficient producer of \fe\ lines, because most of incoming
energy is used to heat the gas and not for ionizing iron. Thus, this
process will only be important in environments where photons are not
as efficiently produced as high energy particles. While detailed
calculations are beyond the scope of this paper, non-thermal radio
sources or jet-like flows, where the bulk of the energy may be
transported by relativistic particles, may be areas where particle
impact is the dominant source of ionization. The shock front in a gamma-ray burst is
another possible area of application if there are enough high energy
baryons in the relativistic outflow.

\section{Summary}
\label{sect:summ}
 Just like a photon, high energy electrons and protons can eject a
 K-shell electron from iron and produce a K$\alpha$ line photon if
 their kinetic energy is $> 7.112$~\kev. In this paper we have considered the
 possible role of particle impact in the generation of \fe\ emission
 from accreting black holes. Calculations of \fe\ flux from various
 columns of gas were performed when it was bombarded by a non-thermal
 power-law of electrons, or a thermal distribution of electrons or
 protons. This flux was then compared with that produced by a
 power-law of radiation under the assumption that all process
 deposited roughly 1~erg~cm$^{-2}$~s$^{-1}$ of energy into the
 material. It was found that in every circumstance the \fe\ flux
 produced by particle impact was $< 1$\% of that produced by
 photoionization. A more realistic calculation which relaxes the
 optically-thin assumption made here would find an even smaller ratio.

Increasing the electron flux on the accretion disk
in order to produce more K$\alpha$ emission was inconsistent with the
standard disk-corona model for accretion disk emission. In the case of
a disk illuminated by virialized protons, as might be found in a
truncated accretion disk model, we found a very low efficiency for \fe\
production. Therefore, these type of models must consider external
X-ray illumination to explain any observed K$\alpha$ line.

To conclude, we find that, except for isolated regions around magnetic
footpoints, it is unlikely that particle impact can significantly
contribute to the observed X-ray line (or continuum) emission from
accreting black holes. Depending on the efficiency of the unknown
acceleration mechanisms, particles may, however, be a important
heating source for accretion disks.

\acknowledgments

DRB was supported by the Natural Sciences and Engineering
Research Council of Canada. ACF thanks the Royal Society for support.
We thank the anonymous referee for useful comments.

\appendix
\section{Unnormalized Line Fluxes from Particle Impact}

\begin{figure}
\includegraphics[width=0.75\textwidth,angle=-90]{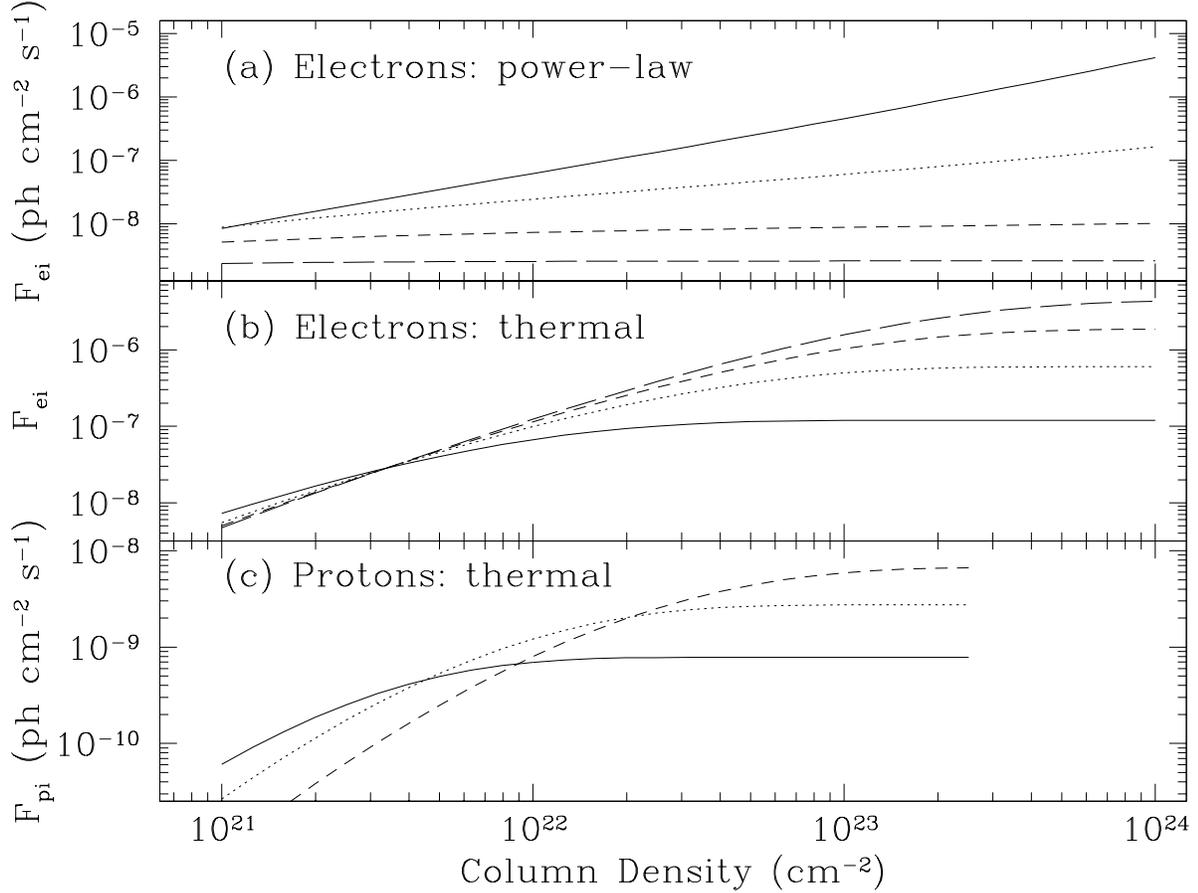}
\caption{Iron K$\alpha$ flux in photons~cm$^{-2}$~s$^{-1}$ produced by
  particle impact as a function of hydrogen column density. A solar
  abundance of iron has been assumed. (a) The line flux produced by
  electron impact when the particles have a power-law spectrum. The
  energy flux between 7.112~\kev\ and 200~\kev\ was
  1~keV~cm$^{-2}$~s$^{-1}$. The line styles are the same as in
  Fig.~\ref{fig:eplaw}. (b) The line flux produced by electron impact
  when the particles have a thermal spectrum. The energy flux between
  7.112~\kev\ and 200~\kev\ was 1~keV~cm$^{-2}$~s$^{-1}$. The line
  styles are the same as in Fig.~\ref{fig:emax}. (c) The line flux
  produced by proton impact when the particles have a thermal
  spectrum. The total energy flux was 1~keV~cm$^{-2}$~s$^{-1}$. The
  line styles are the same as in Fig.~\ref{fig:pmax}, and the 50~MeV
  cutoff in the proton ionization cross-section limited the
  calculation to lower column densities.}
\label{fig:flx}
\end{figure}

In the interests of generality, we show in Figure~\ref{fig:flx} the
unnormalized \fe\ line fluxes (in photons~cm$^{-2}$~s$^{-1}$) obtained
from equations~\ref{eq:flx-ei} and \ref{eq:flx-pi} for the power-law
and thermal spectral shapes used in \S~\ref{sect:res}. The standard
values of $\omega=0.342$ and $\beta=0.822$ from \citet{bam72} were
used, and the solar Fe abundance of $A_{\mathrm{Fe}}=2.82\times
10^{-5}$ from \citet{hol01} was assumed. As the Fe abundance is just a
multiplicative factor, the results can be easily scaled to other
metallicities. Finally, the particle spectra were normalized as in
\S~\ref{sect:res}; that is, the electron spectra has a energy
flux between 7.112~\kev\ and 200~\kev\ of 1~keV~cm$^{-2}$~s$^{-1}$,
while the protons have a total energy flux of
1~keV~cm$^{-2}$~s$^{-1}$. Again, the results in Fig.~\ref{fig:flx} can
be easily scaled to other impacting fluxes.

\clearpage


\end{document}